\documentclass[aps,prl,amsmath,twocolumn,10pt,superscriptaddress,floatfix,nofootinbib]{revtex4-1}
\usepackage{epsfig,amssymb,amsfonts,amsmath,mathtools,bm,color,xcolor,graphicx,braket,adjustbox,esint,upgreek}
\usepackage{multirow}
\usepackage[utf8]{inputenc}
\usepackage{setspace}
\usepackage{dcolumn,kantlipsum}
\allowdisplaybreaks

\newcommand{\be}{\begin{equation}}
\newcommand{\ee}{\end{equation}}
\newcommand{\bea}{\begin{eqnarray}}
\newcommand{\eea}{\end{eqnarray}}
\newcommand{\beas}{\begin{eqnarray*}}
\newcommand{\eeas}{\end{eqnarray*}}

\newcommand{\tr}{\mbox{Tr}}

\def\vec#1{\boldsymbol{#1}}

\newcommand{\pc}{P_c}
\newcommand{\jp}{J/\psi p}
\newcommand{\sigh}{\Sigma_c^{(*)}\bar{D}^{(*)}}
\newcommand{\sigd}{\Sigma_c\bar{D}}
\newcommand{\sigdstar}{\Sigma_c\bar{D}^*}
\newcommand{\sigstard}{\Sigma_c^*\bar{D}}

\newcommand{\bonn}{\affiliation{Helmholtz-Institut f\"ur Strahlen- und Kernphysik and\\ Bethe Center for Theoretical Physics, Universit\"at Bonn, D-53115 Bonn, Germany}}

\newcommand{\fzj}{\affiliation{Institute for Advanced Simulation, Institut f\"ur Kernphysik and\\ J\"ulich Center for Hadron Physics, Forschungszentrum J\"ulich, D-52425 J\"ulich, Germany}}

\newcommand{\itp}{\affiliation{CAS Key Laboratory of Theoretical Physics, Institute of Theoretical Physics,\\ Chinese Academy of Sciences,  Zhong Guan Cun East Street 55, Beijing 100190, China}}

\newcommand{\ucas}{\affiliation{School of Physical Sciences, University of Chinese Academy of Sciences, Beijing 100049, China}}

\newcommand{\itep}{\affiliation{Institute for Theoretical and Experimental Physics  NRC  “Kurchatov Institute”, Moscow 117218, Russia}}

\newcommand{\lebedev}{\affiliation{P.N. Lebedev Physical Institute of the Russian Academy of Sciences, 119991, Leninskiy Prospect 53, Moscow, Russia}}

\newcommand{\tsu}{\affiliation{Tbilisi State University, 0186 Tbilisi, Georgia}}

\newcommand{\murcia}{\affiliation{Departamento de F{\'i}sica, Universidad de Murcia, E-30071 Murcia, Spain}}

\newcommand{\scnu}{\affiliation{{Guangdong Provincial Key Laboratory of Nuclear Science,}\\ Institute of Quantum Matter, South China Normal University, Guangzhou 510006, China}}

\newcommand{\tpcsf}{\affiliation{Theoretical Physics Center for Science Facilities,\\ Institute of High Energy Physics, Chinese Academy of Sciences, Beijing 100049, China}}

\graphicspath{{FiguresV2/}}

\begin{document}

\title{Interpretation of the LHCb $\bm{P_c}$ States as Hadronic Molecules and Hints of a Narrow $\bm{P_c(4380)}$}

\author{Meng-Lin Du}\email{du@hiskp.uni-bonn.de}
\bonn

\author{Vadim Baru}\email{vadim.baru@tp2.rub.de}
\bonn \itep \lebedev

\author{Feng-Kun Guo}\email{fkguo@itp.ac.cn}
\itp \ucas

\author{Christoph~Hanhart}\email{c.hanhart@fz-juelich.de}
\fzj

\author{Ulf-G.~Mei{\ss}ner}\email{meissner@hiskp.uni-bonn.de}
\bonn \fzj \tsu

\author{Jos\'e A. Oller}\email{oller@um.es}
\murcia

\author{Qian Wang}\email{qianwang@m.scnu.edu.cn}
\scnu\tpcsf

\begin{abstract}

Three hidden-charm pentaquark $P_c$ states, $P_c(4312)$, $P_c(4440)$, and $P_c(4457)$ were revealed in the
$\Lambda_b^0\to J/\psi p K^-$ process measured by LHCb using both run I and run II data. Their nature is
under lively discussion, and their quantum numbers have not been determined. We analyze the $\jp$ invariant
mass distributions under the assumption that the crossed-channel effects provide a smooth background.
For the first time, such an analysis is performed employing 
 a coupled-channel formalism with the  scattering potential involving both one-pion exchange as well as short-range operators constrained by heavy quark spin symmetry.
We find that the data can be well described in the hadronic molecular picture, which predicts seven
$\Sigma_c^{(*)}\bar D^{(*)}$ molecular states in two spin multiplets, such that the $P_c(4312)$ is mainly
a $\Sigma_c\bar D$ bound state with $J^P=1/2^-$, while $P_c(4440)$ and $P_c(4457)$ are $\Sigma_c\bar D^*$ bound
states with quantum numbers $3/2^-$ and $1/2^-$, respectively. 
We also show that there is  evidence for a narrow $\Sigma_c^*\bar D$ bound state in the data which we
call $P_c(4380)$, different from the broad one reported by LHCb in 2015. With this state included,
all predicted $\Sigma_c \bar D$, $\Sigma_c^* \bar D$, and $\Sigma_c \bar D^*$ 
hadronic molecules are seen in the data, while the missing three $\Sigma_c^*\bar D^*$ states 
are expected to be found in future runs of the LHC or in photoproduction experiments.

\end{abstract}

\maketitle

{\it Introduction.}--The confinement property of quantum chromodynamics (QCD) in principle allows for the existence of
a large variety of color neutral objects. However, it is not clear yet which configurations are realized in nature.
As a result, searching for multiquark exotic hadrons beyond the conventional quark model has been
one of the central issues in the study of the strong interactions. Tremendous developments have been made
in the new era since 2003 when the $B$ factories discovered the
$D_{s0}^*(2317)$~\cite{Aubert:2003fg} and $X(3872)$~\cite{Choi:2003ue}, whose properties are in
notable contradiction with quark model predictions.
The interest in studying such exotic hadrons was further boosted by the LHCb discovery of the hidden-charm
pentaquarks $P_c(4450)$ and $P_c(4380)$ decaying into $\jp$ in the $\Lambda_b^0\to K^-\jp$ process in
2015~\cite{Aaij:2015tga}.  
The experimental and theoretical efforts are summarized in a number of comprehensive reviews~\cite{Brambilla:2010cs,Chen:2016spr,Chen:2016qju,Dong:2017gaw,Esposito:2016noz,Hosaka:2016pey,Olsen:2017bmm,Guo:2017jvc,Kou:2018nap,Cerri:2018ypt,Liu:2019zoy,Brambilla:2019esw,Guo:2019twa}.
When LHCb updated their measurements with a one-order-of-magnitude
larger data sample~\cite{Aaij:2019vzc}, the narrow $\pc(4450)$ appeared to be split into two narrower structures
$\pc(4440)$ and $\pc(4457)$, and a third narrow peak $\pc(4312)$ showed up. At the same time, the broad $\pc(4380)$
 (whose existence needs to be verified in a complete amplitude analysis that is under way) lost
its significance.
A cornucopia of theoretical interpretations followed these new discoveries, including models of
hadronic molecules~\cite{Chen:2019bip,Chen:2019asm,Guo:2019fdo,Liu:2019tjn,He:2019ify,Guo:2019kdc,Shimizu:2019ptd,Xiao:2019mst,Xiao:2019aya,Wang:2019nwt,Meng:2019ilv,Wu:2019adv,Xiao:2019gjd,Voloshin:2019aut,Sakai:2019qph,Wang:2019hyc,Yamaguchi:2019seo,Liu:2019zvb,Lin:2019qiv,Wang:2019ato,Gutsche:2019mkg,Burns:2019iih}, compact pentaquark states~\cite{Ali:2019npk,Zhu:2019iwm,Wang:2019got,Giron:2019bcs,Cheng:2019obk,Stancu:2019qga}, and hadrocharmonia~\cite{Eides:2019tgv}.
An amplitude analysis was performed in Ref.~\cite{Fernandez-Ramirez:2019koa} focusing on the $\pc(4312)$
which was suggested to be a virtual state.
Among the explanations, the hadronic molecular model stands out as it explains all of the three narrow
$\pc$ states simultaneously as $\sigd$ [for $\pc(4312)$ and $\sigdstar$ [for $\pc(4440)$ and
$\pc(4457)$] bound states, see, e.g., Refs.~\cite{Liu:2019tjn,Xiao:2019aya,Sakai:2019qph}, employing the approximate
heavy quark spin symmetry (HQSS) of QCD.
However, the model predicts in addition four more states, including one $\sigstard$ state at around $4.37$
to $4.38$~GeV and three states slightly below the $\Sigma_c^*\bar D^*$ threshold. These seven $\pc$ states
are contained in two heavy quark spin multiplets, labeled as $j_{\ell}^P$, with $j_{\ell}^{}$ and $P$ the total angular
momentum of the light degrees of freedom and parity, respectively: three with $j_{\ell}^P=\frac12^-$ and
four with $j_{\ell}^P=\frac32^-$. States in these two multiplets with the same spin mix
because the $\Sigma_c (\bar D)$ is not degenerate with the $\Sigma_c^* (\bar D^*)$.
While only three of the states correspond to the ones reported by LHCb, it is crucial to check whether the
existence of the whole two multiplets is consistent with the $\jp$ distribution measured by LHCb.
This is the question addressed in this Letter:
by constructing coupled-channel amplitudes analogous to those used in the analysis of the $Z_b$
states~\cite{Hanhart:2015cua,Guo:2016bjq,Wang:2018jlv,Baru:2019xnh}, we show that the observed
$\jp$ invariant mass distribution can be well described in the hadronic molecular scenario with
seven $\sigh$ molecules, where the $P_c(4312)$ and $P_c(4440)/P_c(4457)$ are $\sigd$ and $\sigdstar$
bound states, respectively. For the first time, we point out a clear signal from data for the existence of a narrow $P_c(4380)$ as a $\sigstard$ bound state. 
The three predicted $\Sigma_c^*\bar D^*$ states still await discovery. 
In addition, we emphasize that the inclusion of the one-pion exchange (OPE)  allows one to single out a unique solution corresponding to the global minimum instead of  two equivalent solutions present in the 
pure contact case.

{\it Framework.}-- In order to describe the measured $\jp$ distribution, we construct coupled-channel amplitudes considering
all the $\sigh$ channels (called elastic channels~\cite{Hanhart:2015cua,Guo:2016bjq},
since their thresholds are close to the $P_c$ masses) and the $\jp$ channels (inelastic channels).
HQSS is used to relate all the $\sigh$ channels~\cite{Xiao:2013yca,Liu:2018zzu,Liu:2019tjn,Sakai:2019qph},
and their couplings to the $\jp$~\cite{Voloshin:2019aut,Sakai:2019qph}. We also allow for additional inelastic
channels not included explicitly in the amplitudes.
To this end, we expand the two-particle states in the basis of HQSS eigenstates  $|s_Q\otimes j_{\ell}\rangle$,
with $s_Q$ and $j_{\ell}$ representing the total spin of the heavy quarks and total angular momentum
of light degrees of freedom, respectively.  In this notation, the $\Sigma^{(*)}$ and $\bar D^{(*)}$ spin multiplets
are $|\frac12\otimes 1\rangle$ and $|\frac12\otimes\frac{1}{2}\rangle$, respectively. 
One can rewrite the $S$-wave $\Sigma_c^{(\ast)}\bar{D}^{(\ast)}$ systems in terms of
$|s_Q\otimes j_{\ell}\rangle$ as~\cite{Xiao:2013yca,Sakai:2019qph}
\begin{eqnarray}
\left(\begin{array}{c}
|\Sigma_{c}\bar{D}\rangle\\
|\Sigma_{c}\bar{D}^{*}\rangle\\
|\Sigma_{c}^{*}\bar{D}^{*}\rangle
\end{array}\right)_{\frac{1}{2}} = \left(\begin{array}{ccc}
\frac{1}{2} & \frac{-1}{2\sqrt{3}} & \sqrt{\frac{2}{3}}\\
\frac{-1}{2\sqrt{3}} & \frac{5}{6} & \frac{\sqrt{2}}{3}\\
\sqrt{\frac{2}{3}} & \frac{\sqrt{2}}{3} & -\frac{1}{3}
\end{array}\right)\left(\begin{array}{c}
|0\otimes\frac{1}{2}\rangle\\
|1\otimes\frac{1}{2}\rangle\\
|1\otimes\frac{3}{2}\rangle
\end{array}\right),~~\label{eq:HL1} 
\end{eqnarray}
\begin{eqnarray}
\left(\begin{array}{c}
|\Sigma_{c}\bar{D}^{*}\rangle\\
|\Sigma_{c}^{*}\bar{D}\rangle\\
|\Sigma_{c}^{*}\bar{D}^{*}\rangle
\end{array}\right)_{\frac{3}{2}} = \left(\begin{array}{ccc}
\frac{-1}{\sqrt{3}} & \frac{1}{3} & \frac{\sqrt{5}}{3}\\
\frac{1}{2} & \frac{-1}{\sqrt{3}} & \frac{1}{2}\sqrt{\frac{5}{3}}\\
\frac{1}{2}\sqrt{\frac{5}{3}} & \frac{\sqrt{5}}{3} & \frac{1}{6}
\end{array}\right)\left(\begin{array}{c}
|0\otimes\frac{3}{2}\rangle\\
|1\otimes\frac{1}{2}\rangle\\
|1\otimes\frac{3}{2}\rangle
\end{array}\right),~~\label{eq:HL2}
\end{eqnarray}
\begin{eqnarray}
|\Sigma_{c}^{*}\bar{D}^{*}\rangle_{\frac{5}{2}}=|1\otimes\frac{3}{2}\rangle,\label{eq:HL3}
\end{eqnarray}
where the subscripts on the left-hand side represent the total angular momentum $J=\frac{1}{2},\frac{3}{2}$,
and $\frac{5}{2}$. The rotation matrices in Eqs.~\eqref{eq:HL1}--\eqref{eq:HL3} will be denoted as $R^{J}$ in the following. 
One can obtain the contact terms for $S$-wave interactions of the elastic
channels in terms of two independent matrix elements, 
\begin{equation}  \nonumber
  C_{\frac{1}{2}}\equiv\langle s_Q\otimes\frac{1}{2}|\hat{\mathcal{H}}_{I}|s_Q\otimes\frac{1}{2}\rangle,\quad C_{\frac{3}{2}}\equiv\langle s_Q\otimes\frac{3}{2}|\hat{\mathcal{H}}_{I}|s_Q\otimes\frac{3}{2}\rangle,
  \label{eq:contact}
\end{equation}
with $\hat{\mathcal{H}}_I$ the effective Hamiltonian respecting HQSS.  
In the heavy quark limit, the contact interactions defined above are independent of $s_Q=0, 1$.
Since we work to leading order, the above matrix elements are constants. 
In particular, the $D$-wave $\Sigma_c^{(\ast)}\bar{D}^{(\ast)}$ contact terms, which turned out to be necessary in the study
of the $Z_b$ states~\cite{Baru:2019xnh},  will be neglected since data in the inelastic channels are insensitive to such operators. 
The contact terms in the particle basis
are 
\begin{equation}
C^J_{\alpha\beta} = \sum_{n_J}R^J_{\alpha\, n_J}\ C_{j_\ell(n_J)} \ \left(R^J\right)^T_{n_J\, \beta} \ ,
\end{equation}
where $j_\ell(n_J)$ denotes the light-quark spin of the $n$th channel for
a given $J$ multiplet.

The OPE potential can be obtained using the effective Lagrangian for the axial
coupling of the pions to the charmed mesons and baryons~\cite{Wise:1992hn,Yan:1992gz} 
\bea
\mathcal{L}&=&\frac{g}{4}\left\langle\vec\sigma\cdot \vec{u}_{ab}\bar{H}_{b}\bar{H}_{a}^{\dagger}\right\rangle
-i g_{1}\epsilon^{ijk}\,\tr\left[{S}_i^\dag u_j S_k \right] , 
\label{lag:1}
\eea
where $\langle.\rangle$ and $\tr[.]$ denote traces in the spinor and isospin spaces, respectively, $\vec\sigma$
represents the Pauli matrices,  $S_i$ and $\bar{H}$ are the heavy quark spin doublets for ground states
$(\Sigma_c,\Sigma_c^\ast)$ and $(\bar{D},\bar{D}^\ast)$ \cite{Hu:2005gf},
\bea
\vec{S} = \frac{1}{\sqrt{3}}\vec{\sigma}\Sigma_c + \vec{\Sigma_c^\ast},\quad \bar{H}=-\bar{D}
+\vec{\sigma}\cdot\vec{\bar{D}^\ast},
\eea 
$\vec{u}=-i\nabla\Phi/F_{\pi}+\mathcal{O}(\Phi^{3})$, $\Phi = \vec\tau \cdot \vec\pi$ with $\vec\tau$ and $\vec\pi$
the Pauli matrices in the isospin space and the pion fields, in order,
and $F_\pi=92.1~\mathrm{MeV}$ is the pion decay constant.
From the measured width of $D^{*+}\to D^0\pi^+$
one gets $g=0.57$~\cite{Tanabashi:2018oca}, and the coupling 
$g_1=0.42$ is taken from the lattice QCD calculation~\cite{Detmold:2012ge}. The OPE contributes to both $S$ and
$D$ waves and can be important for describing the line shapes around thresholds~\cite{Baru:2017gwo,Wang:2018jlv,Baru:2019xnh}.

Also the transitions between the elastic and inelastic channels can be related via HQSS.
While the $|1\otimes\frac{1}{2}\rangle$ component in Eqs.~\eqref{eq:HL1}--\eqref{eq:HL3} couples to
$J/\psi p$ in the $S$ wave in the heavy quark limit, the $|1\otimes\frac{3}{2}\rangle$ only couples
to $J/\psi p$ in the $D$ wave. We introduce two coupling strengths,
$$g_S \equiv \langle 1\otimes \frac{1}{2}|\hat{\mathcal{H}}_I|J/\psi p\rangle_S, \quad
g_Dk^2 \equiv \langle 1\otimes \frac{3}{2}|\hat{\mathcal{H}}_I|J/\psi p\rangle_D,$$ 
where $k$ is the magnitude of the $J/\psi$ three-momentum in the  c.m. frame of $J/\psi p$. 
Then, the transition vertices $\mathcal{V}_{\alpha i}^J$ between the $\alpha$th elastic and $i$th inelastic channel,
with $i=1, 2$ denoting the $S$-wave and $D$-wave $J/\psi p$, respectively, can be easily obtained by
virtue of the decompositions in Eqs.~\eqref{eq:HL1}--\eqref{eq:HL3} as
\begin{align}
&\mathcal{V}^J_{\alpha 1} = g_SR_{\alpha 2}^J, \quad &\mathcal{V}^J_{\alpha 2}(k)&=g_D k^2R_{\alpha 3}^J,
  \quad &J &=\frac12,\,\frac32, \nonumber\\
&\mathcal{V}^J_{\alpha 1} = 0, \quad &\mathcal{V}^J_{\alpha 2}(k)&=g_D k^2, \quad &J &=\frac52.
\end{align}
The direct $\jp$ interaction is Okubo-Zweig-Iizuka suppressed and found very weak in
a recent lattice QCD study~\cite{Skerbis:2018lew}. 
Thus, the inelastic $\jp$ channel is only included through its coupling to the elastic 
channels~\cite{Hanhart:2015cua,Albaladejo:2015lob,Guo:2016bjq,Wang:2018jlv,Baru:2019xnh}.
While the real part of its contribution can be absorbed by redefining the contact
terms $C_{\alpha\beta}^J$~\cite{Baru:2019xnh}, the imaginary part cannot.
We thus introduce
\begin{eqnarray}
V^J_{J/\psi p,\alpha\beta}(E)=-\frac{i}{2\pi E} \sum_{j=1}^2 m_{J/\psi}m_p \mathcal{V}^J_{\alpha j}
\mathcal{V}^J_{\beta j}\,k
\label{eq:jp}
\end{eqnarray} 
into the effective elastic potential.
It is expected that, in addition to the $J/\psi p$ channels, there are more inelastic channels,
most prominently $\Lambda_c \bar D^{(*)}$ and $\eta_c p$~\cite{Wu:2012md,Lin:2017mtz,Voloshin:2019aut}. While the latter is
connected to the $J/\psi p$ channels via HQSS, the former is not and thus
we are obliged to parametrize especially those via an additional imaginary part of
the two contact terms.
This introduces two more parameters. Thus the scattering problem
contains in total 6 parameters and the full effective potential
for the elastic channels can be written as
\begin{eqnarray}
  V^J(E,p,q)=C^J+V_{J/\psi p}^J(E)+V^J_\mathrm{OPE}(E,p,q) \ ,
  \label{eq:vfull}
\end{eqnarray}
with $q$ and $p$ for the incoming and outgoing relative momenta of the corresponding channels, respectively,
and $E$ for the total energy. The explicit expressions for the matrix $V^J_\mathrm{OPE}(E,p,q)$ 
are provided in Ref.~\cite{supp}.

For the weak production amplitude for $\Lambda_b\to K^-\sigh$ we only consider the elastic
$\Sigma_c^{(\ast)}\bar{D}^{(\ast)}$ channels produced in an $S$ wave, since the energy region of interest is close to the $\sigh$ thresholds.
Then the weak production matrix elements  may be expressed as $\mathcal{F}^J_n = \langle \Lambda_b |\hat{\mathcal{H}}_W|K^-(s_Q\otimes j_{\ell})_n^J\rangle$,
where $(s_Q\otimes j_{\ell})_n^J$ refers to the $n$th state in the $|s_Q\otimes j_{\ell}\rangle$
basis in Eqs.~\eqref{eq:HL1}--\eqref{eq:HL3}.
The production contact term for the $\alpha$th elastic channel  for a  given $J$ then reads
\bea
P_\alpha^J = \sum_n R^J_{\alpha n} \mathcal{F}_n^J~. 
\label{eq:production}
\eea 
In total, there are additional seven parameters $\mathcal{F}_n^J$.

With the above ingredients, one can obtain the production amplitude, $U^J_\alpha$, for the $\alpha^{\rm th}$ 
elastic channel by solving the following Lippmann-Schwinger equations (LSEs), 
\begin{equation}
U^J_\alpha(E,p) {=} P^J_\alpha {-}\! \sum_\beta \!\! \int \!\! \frac{d^3\vec{q}}{(2\pi)^3}V^J_{\alpha\beta}(E,p,q)G_\beta(E,q) U^J_\beta(E,q),
\label{eq:lse}
\end{equation}
and for the $i$th $J/\psi p$ inelastic channel $U^J_i$ via
\begin{eqnarray}
U^J_i(E,k){=} {-}\! \sum_\beta \!\! \int \!\! \frac{d^3 \vec{q}}{(2\pi)^3}\mathcal{V}^J_{i\beta}(k) G_\beta(E,q) U^J_\beta(E,q).
\label{eq:inelprod}
\end{eqnarray}
The two-body propagator is
 \begin{eqnarray}
 G_\beta(E,q)=\frac{2\mu_\beta}{q^2-p_\beta^2-i\epsilon},\quad p_\beta^2\equiv2\mu_\beta(E-m_\mathrm{th}^\beta),
 \end{eqnarray}
 with $\mu_\beta$ and $m_\mathrm{th}^\beta$ the reduced mass and the threshold of the $\beta$th elastic
 channel. The $\Sigma_c^{(*)}$ widths of 1.86~MeV (15~MeV)~\cite{Tanabashi:2018oca} are accounted for using a complex mass
 $m-i\Gamma/2$ in $m_\mathrm{th}^\beta$.
 The LSE is regularized using a hard cutoff, varied in the range from  1  to 1.5~GeV. 
 Since the results barely depend on its value (effects
 of the cutoff variation can be largely absorbed into the refitted contact terms), the final results will be presented for the cutoff of 1~GeV. 
 The equations given are unitary as long as the additional imaginary parts of the contact terms
 are omitted. Unitarity can be restored once data on the $\Lambda_c\bar D^{(*)}$ channels are available, and we checked that
 introducing the $\Lambda_c\bar D^{*}$ channel in the way of Eq.~\eqref{eq:jp} did not produce a sizable difference.

In order to fit the $J/\psi p$ invariant mass distribution, an incoherent smooth background is used to model
possible contributions from misidentified non-$\Lambda_b^0$ events, the $\Lambda^\ast$ resonances coupled
to  $p K^-$, and possibly additional broad $P_c^+$ structures. Here we use the form 
\begin{equation}
f_\mathrm{bgd}(E)=b_0+b_1E^2+b_2E^4+\left|\frac{g_r}{m^2-E^2-i\,\Gamma E}\right|^2,
\label{eq:back}
\end{equation}
which contains the parameters $b_0$, $b_1$, $b_2$, $g_r$, $m$, and $\Gamma$. The backgrounds used in the experimental
analysis~\cite{Aaij:2019vzc} are also considered, and the results are similar, which will be included in the uncertainties. 
We perform fits employing the potential of Eq.~\eqref{eq:vfull} either omitting (scheme~I) or including (scheme~II) $V_\text{OPE}^J$.

\begin{table*}
\caption{The names of the states, their quantum numbers
  found from the fits within scheme II, the pole positions (on the sheets close to the physical one),
  the dominant channels (DCs) and their thresholds,
the dimensionless couplings of the resonances in the DCs (from the $T$-matrix residues and defined as $G_{\text{DC}}$), and
the resonance couplings to the source derived from the residues, which are normalized by the event numbers and thus only the relative values are meaningful.
The uncertainties given are from taking different backgrounds, the uncertainties
from the fit for a given background are negligible. 
The $P_c(4380)$ in boldface is the new state we advocate in this work.
}
\begin{ruledtabular}
\begin{tabular}{|l|c c c c c|}
Scheme II & $J^{P}$ & Pole {[}MeV{]} & DC (threshold [MeV]) & $G_{\text{DC}}$ & Production \tabularnewline
\hline 
$P_{c}(4312)$ & $\frac{1}{2}^{-}$  &  $4314(2)-5(2)i$  & $\Sigma_{c}\bar{D}\; (4321.6)$ & $2.86(12)-0.44(24)i$  & $636(73)-98(53)i$\tabularnewline
$\mathbf{P_{c}(4380)}$ & $\frac{3}{2}^{-}$ & $4378(2)-13(3)i$  & $\Sigma_{c}^{*}\bar{D}$ (4386.2) & $3.00(12)-0.49(27)i$ & $618(373)-181(95)i$ \tabularnewline
$P_{c}(4440)$ & $\frac{3}{2}^{-}$  & $4441(2)-11(3)i$ & $\Sigma_{c}\bar{D}^{*}$ (4462.1) & $3.91(11)-0.62(19)i$ & $999(140)-15(18)i$ \tabularnewline
$P_{c}(4457)$ & $\frac{1}{2}^{-}$  & $4459(2)-4(1)i$ & $\Sigma_{c}\bar{D}^{*}$ (4462.1) & $2.09(17)-0.46(18)i$ & $-918(68)+159(78)i$ \tabularnewline
$P_{c}$ & $\frac{1}{2}^{-}$  & $4524(2)-9(1)i$ & $\Sigma_{c}^{*}\bar{D}^{*}$ (4526.7) & $1.90(23)-0.28(21)i$ & $-228(384)+22(23)i$ \tabularnewline
$P_{c}$ & $\frac{3}{2}^{-}$  & $4518(2)-11(2)i$ & $\Sigma_{c}^{*}\bar{D}^{*}$ (4526.7)& $2.83(16)-0.43(18)i$ & $-156(517)-58(43)i$ \tabularnewline
$P_{c}$ & $\frac{5}{2}^{-}$  & $4498(5)-35(17)i$ & $\Sigma_{c}^{*}\bar{D}^{*}$ (4526.7)& $4.66(55)-1.12(32)i$ & $-393(620)-2(26)i$   
\end{tabular}
\end{ruledtabular} 
\label{tab:pole}
\end{table*}

\begin{figure*}[!htb]
 \centering
  \includegraphics[width=\textwidth]{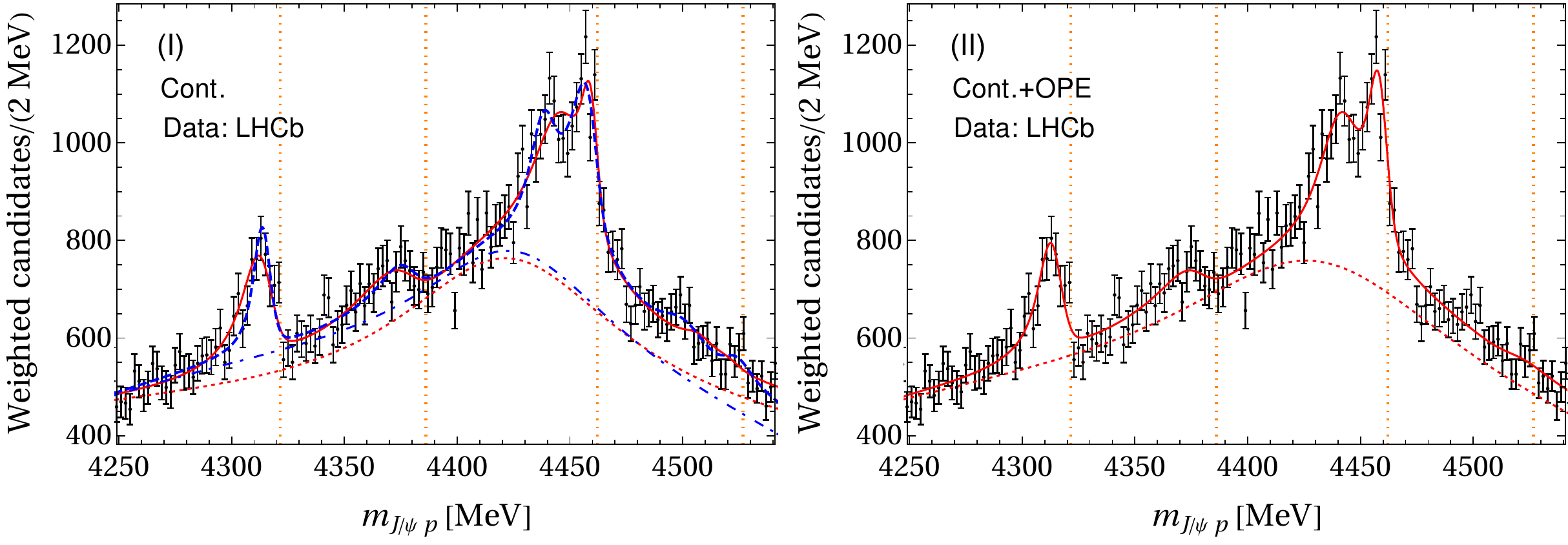}
  \caption{Left panel: the fitted invariant mass distributions versus the weighted experimental data~\cite{Aaij:2019vzc}
    for both solution~$A$ (blue dashed curves) and solution~$B$ (red solid ones) of scheme~I. The corresponding
    backgrounds are shown as blue dash-dotted and red dotted curves, respectively. Right panel: the
    best fit for scheme~II with the dotted curve representing the background.
    The vertical dashed lines in both panels from left to right are the $\Sigma_c\bar{D}$,
    $\Sigma_c^*\bar{D}$, $\Sigma_c\bar{D}^*$, and $\Sigma_c^*\bar{D}^*$ thresholds, respectively. 
    }
  \label{fig:FitBOPE}
\end{figure*}

{\it Results and discussions.}--In this analysis, we do not consider isospin symmetry breaking effects
which can be important to give rise to isospin-breaking decay modes~\cite{Burns:2015dwa,Guo:2019fdo} but
have little effect on the description of the line shapes in the isospin-conserving $\jp$ channel.  Convolution with the experimental
energy resolution is considered. For scheme~I we find two solutions, denoted as $A$ and $B$, describing the data almost equally well
(with $\chi^2/\text{d.o.f.}=1.01$ and $1.03$, respectively). The corresponding best fits are shown in the left panel
of Fig.~\ref{fig:FitBOPE}. 
The two solutions produce different values of the parameters, in particular $C_\frac12$ and $C_\frac32$
(in fact, the values of $C_\frac12$ and $C_\frac32$ in solution~A are very close to those of $C_\frac32$ and
$C_\frac12$ in solution~B, respectively), and thus give different pole locations. However,
both solutions give seven poles in the $\Sigma_c^{(\ast)}\bar{D}^{(\ast)}$ scattering
amplitudes, i.e., seven $\pc$ states: three with $J=1/2$, three with $J=3/2$ and one with $J=5/2$.  The masses of the generated
$\pc$ states in solutions~$A$ and $B$ are close to those of scenarios~$A$ and $B$ for the corresponding
quantum numbers in Refs.~\cite{Liu:2019tjn,Sakai:2019qph}, respectively, though the widths are larger due to the inelastic channels and
 the $\Sigma_c^{(*)}$ widths, and thus are not shown here. 

In  both solutions, among the seven poles, the lowest one corresponds to the $P_c(4312)$ with
$\frac12^-$, and is a $\Sigma_c\bar{D}$ bound state: it is located in the second Riemann sheet
(of the $\jp$ channel), and would become a real bound state pole in the first Riemann sheet if the $\jp$
channel were switched off. This is different from the virtual state scenario of $\Sigma_c\bar{D}$ in
Ref.~\cite{Fernandez-Ramirez:2019koa} which only fits to data around the $\sigd$ threshold. 

There are two $\Sigma_c\bar{D}^*$ bound states with quantum numbers $\frac{1}{2}^-$ and $\frac{3}{2}^-$,
corresponding to the $P_c(4440)$ and the $P_c(4457)$, respectively, in solution~$A$ and interchanged in solution~$B$. 
The mass pattern of the three $P_c$ states dominated by the $\Sigma_c^*\bar{D}^*$ channel is analogous
to that of the $\Sigma_c\bar{D}^*$ channel,  i.e., $m_{\frac{1}{2}^-}<m_{\frac{3}{2}^-}<m_{\frac{5}{2}^-}$ 
for solution~$A$ and the opposite for solution~$B$.
In both solutions, there is a narrow pole around $4.38~\mathrm{GeV}$ whose dominant component is $\sigstard$ (see also Ref.~\cite{Xiao:2019aya}).
This means that  HQSS  requires the existence of a $P_c(4380)$ resonance which, however, is narrow and
thus different from the broad resonance reported by LHCb in 2015~\cite{Aaij:2015tga}.

In scheme~II the OPE is included.
The importance of its  tensor force is well known
for the nucleon-nucleon interaction. It leads to the
mixing between $S$ and $D$ waves
and can have a sizable impact on the line shape between thresholds~\cite{Baru:2017gwo,Wang:2018jlv,Baru:2019xnh}.
Unlike in scheme~I, once the full OPE is included, there is only one solution corresponding to the best fit
with $\chi^2/\text{d.o.f.}=0.98$, shown in the right panel of Fig.~\ref{fig:FitBOPE}. 
It leads to poles presented in Table~\ref{tab:pole} which are similar to those in solution~$B$
of scheme~I (see also Refs.~\cite{Valderrama:2019chc,Liu:2019zvb}).
The pole positions, the dominant channels (DCs) having the largest effective couplings (derived form residues
of the $T$ matrix), and effective couplings of these are listed in Table~\ref{tab:pole}.
The results are insensitive to the form of the background, and the effects of using different backgrounds
give the errors in Table~\ref{tab:pole} (the statistical errors propagating from the data are much
smaller).

One sees that the $\pc(4312)$ couples dominantly to $\sigd$ with $J^P=\frac12^-$, and both the $\pc(4440)$
and $\pc(4457)$ couple dominantly to the $\sigdstar$ with quantum numbers $\frac32^-$ and $\frac12^-$,
respectively. They are all bound state poles and should be understood as hadronic molecules of the corresponding
channels~\cite{Guo:2017jvc}. Comparing the fits from scheme~I and scheme~II, one sees that the OPE helps
making the dip between the $\pc(4440)$ and $\pc(4457)$ more evident.
In both schemes there is a narrow $\pc(4380)$ located at the right position where the data show a peak,
though less prominent than those of the well-known three $P_c$ states. Its existence and properties are a consequence of HQSS
in the hadronic molecular picture. We have checked that it persists no matter whether or not the data
around 4.38~GeV are included in the fit. 
 The absolute value of the production strength of the $P_c(4380)$ shown in the last column of the table differs from zero by $1.7\sigma$ which might
 be taken as a measure of the significance of this resonance in the data.
The three $\sigstard^*$ molecules, which are expected to
exist~\cite{Xiao:2013yca,Liu:2018zzu,Liu:2019tjn,Sakai:2019qph,Wang:2019ato}, do not have any unambiguous signal
in the data.   Their production strengths related to the residues of the production amplitude $U^J_\alpha$ at the pole
 suggest that they are less strongly produced in $\Lambda_b$ decays. 
One possible reason could be that the production of the three most pronounced $\pc$ structures gets
enhanced by nearby triangle singularities discussed in Refs.~\cite{Guo:2015umn,Liu:2015fea,Bayar:2016ftu,Aaij:2019vzc}.

{\it Summary and outlook.}--In summary, we investigated for the first time whether the appealing hadronic molecular model
for the observed $\pc$ states is consistent with the LHCb data. A coupled-channel formalism  is used to analyze
the $\jp$ invariant mass distribution, which contains much more information than the extracted pentaquark masses only. 
The relevant effective potential   constructed based on HQSS involves all  transitions between the elastic $\sigh$
channels, transitions from the elastic to the $S$- and $D$-wave $\jp$ inelastic channels  as well as the coupling
to additional effective inelastic channels. 
We find that the data can be well described. In addition to the three established states, in
our analysis a narrow $\pc(4380)$ state, identified as a $\frac32^-$ $\sigstard$ molecule,
with its properties fixed by HQSS, shows up as a clear signal in the data.
The three $\sigstard^*$ bound states with masses from around 4.49 to 4.52~GeV are almost invisible
because of their relatively  low production rates in the $\Lambda_b$ decays.
We expect that they can be resolved in the forthcoming data to be collected at the LHC run-3 period
or other production processes, such as the $J/\psi$ photoproduction~\cite{Ali:2019lzf,Wang:2015jsa,Kubarovsky:2015aaa,Karliner:2015voa,Blin:2016dlf,Lin:2017mtz,Cao:2019kst,Wang:2019krd,Winney:2019edt,Rossi:2019szt}.
It should be stressed that, if the $P_c$ states indeed are hadronic molecules, they have to show
up as prominent structures also in the elastic channels, and data for those would therefore be extremely
valuable. To further refine our approach, data are needed in the $\Lambda_c\bar D^{(*)}$ 
as well as in the $\eta_c p$ channels. The latter would provide additional information on the amount of spin symmetry violation 
in the system. All these studies 
will shed important light on our understanding of how QCD forms hadrons.

\medskip

\begin{acknowledgements}
We are grateful to Marek Karliner,  Misha Mikhasenko, Sebastian Neubert, Juan Nieves, and Qiang Zhao for useful comments and discussions.
This work is supported in part by the National Natural Science Foundation of China (NSFC) and  the
Deutsche Forschungsgemeinschaft (DFG) through the funds provided to the Sino-German Collaborative Research
Center ``Symmetries and the Emergence of Structure in QCD"  (NSFC Grant No. 11621131001 and DFG Grant No. TRR110),
by the NSFC under Grants No. 11835015, No. 11947302, and  No.  11961141012, by the Chinese Academy of Sciences (CAS) under Grants
No. QYZDB-SSW-SYS013 and No. XDPB09, by
the CAS Center for Excellence in Particle Physics (CCEPP), and by the Munich Institute for Astro- and Particle Physics (MIAPP) of the DFG cluster of excellence ``Origin and Structure of the Universe.'' Q.W. is also supported by the
Thousand Talents Plan for Young Professionals and research startup funding at SCNU.
The work of U.G.M. is further supported by the CAS President's International Fellowship
Initiative (PIFI) (Grant No. 2018DM0034) and by the VolkswagenStiftung (Grant No. 93562).
The work of V.B. is supported by the Russian Science Foundation (Grant No. 18-12-00226).
J.A.O. would like to acknowledge partial financial support  by the MINECO (Spain) and FEDER grant FPA2016-77313-P.
\end{acknowledgements}

\begin{onecolumngrid}
\appendix
\newpage

\begin{center}
  \textbf{\large\normalfont\bfseries\boldmath Interpretation of the LHCb $\bm{P_c}$ States as Hadronic Molecules and Hints of a Narrow $\bm{P_c(4380)}$} \\
\vspace{0.05in}
{ \it \large Supplementary Material}\\
\vspace{0.05in}
\end{center}

\section{Contact Potentials}

The decompositions of the $S$-wave $\Sigma_c^{(*)}\bar{D}^{(*)}$ two-body systems
in terms of the heavy and light degrees of freedom $\left|s_Q\otimes j_{\ell}\right\rangle$ are
\begin{align*}
\left|\Sigma_{c}\bar{D}\right\rangle_{\frac{1}{2}^{-}} & =\frac{1}{2}\left|0\otimes\frac{1}{2}\right\rangle-\frac{1}{2\sqrt{3}}\left|1\otimes\frac{1}{2}\right\rangle+\sqrt{\frac{2}{3}}\left|1\otimes\frac{3}{2}\right\rangle,\\
\left|\Sigma_{c}\bar{D}^{*}\right\rangle_{\frac{1}{2}^{-}} & =-\frac{1}{2\sqrt{3}}\left|0\otimes\frac{1}{2}\right\rangle+\frac{5}{6}\left|1\otimes\frac{1}{2}\right\rangle+\frac{\sqrt{2}}{3}\left|1\otimes\frac{3}{2}\right\rangle,\\
\left|\Sigma_{c}^*\bar{D}^{*}\right\rangle_{\frac{1}{2}^{-}} & =\sqrt{\frac{2}{3}}\left|0\otimes\frac{1}{2}\right\rangle+\frac{\sqrt{2}}{3}\left|1\otimes\frac{1}{2}\right\rangle-\frac{1}{3}\left|1\otimes\frac{3}{2}\right\rangle,
\end{align*}
\begin{align*}
\left|\Sigma_{c}\bar{D}^{*}\right\rangle_{\frac{3}{2}^{-}} &=-\frac{1}{\sqrt{3}}\left|0\otimes\frac{3}{2}\right\rangle+\frac{1}{3}\left|1\otimes\frac{1}{2}\right\rangle+\frac{\sqrt{5}}{3}\left|1\otimes\frac{3}{2}\right\rangle,\\
\left|\Sigma_{c}^*\bar{D}\right\rangle_{\frac{3}{2}^{-}} & =\frac{1}{2}\left|0\otimes\frac{3}{2}\right\rangle-\frac{1}{\sqrt{3}}\left|1\otimes\frac{1}{2}\right\rangle+\frac{1}{2}\sqrt{\frac{5}{3}}\left|1\otimes\frac{3}{2}\right\rangle,\\
\mbox{\ensuremath{}}\left|\Sigma_{c}^*\bar{D}^{*}\right\rangle_{\frac{3}{2}^{-}} & =\frac{1}{2}\sqrt{\frac{5}{3}}\left|0\otimes\frac{3}{2}\right\rangle+\frac{1}{3}\sqrt{5}\left|1\otimes\frac{1}{2}\right\rangle+\frac{1}{6}\left|1\otimes\frac{3}{2}\right\rangle,
\end{align*}
\begin{align*}
\left|\Sigma_{c}^*\bar{D}^{*}\right\rangle_{\frac{5}{2}^{-}} & =\left|1\otimes\frac{3}{2}\right\rangle,
\end{align*}
where the subindex is the total angular momentum, and  $s_Q$ and $j_{\ell}$ represent the total spin of the heavy quarks and total angular momentum
of light degrees of freedom, respectively. 

By defining the contact potentials
\begin{equation}  \nonumber
  C_{\frac{1}{2}}\equiv\left\langle s_Q\otimes\frac{1}{2}\right|\hat{\mathcal{H}}_{I}\left|s_Q\otimes\frac{1}{2}\right\rangle,\quad C_{\frac{3}{2}}\equiv\left\langle s_Q\otimes\frac{3}{2}\right|\hat{\mathcal{H}}_{I}\left|s_Q\otimes\frac{3}{2}\right\rangle,
  \label{eq:contact}
\end{equation}
 one can obtain the effective contact potentials
as
\bea
V_{\frac{1}{2}^{-}}^{C}=\left(\begin{array}{ccc}
\frac{1}{3}C_{\frac{1}{2}}+\frac{2}{3}C_{\frac{3}{2}} & -\frac{2}{3\sqrt{3}}C_{\frac{1}{2}}+\frac{2}{3\sqrt{3}}C_{\frac{3}{2}} & \frac{1}{3}\sqrt{\frac{2}{3}}C_{\frac{1}{2}}-\frac{1}{3}\sqrt{\frac{2}{3}}C_{\frac{3}{2}}\\
-\frac{2}{3\sqrt{3}}C_{\frac{1}{2}}+\frac{2}{3\sqrt{3}}C_{\frac{3}{2}} & \frac{7}{9}C_{\frac{1}{2}}+\frac{2}{9}C_{\frac{3}{2}} & \frac{\sqrt{2}}{9}C_{\frac{1}{2}}-\frac{\sqrt{2}}{9}C_{\frac{3}{2}}\\
\frac{1}{3}\sqrt{\frac{2}{3}}C_{\frac{1}{2}}-\frac{1}{3}\sqrt{\frac{2}{3}}C_{\frac{3}{2}} & \frac{\sqrt{2}}{9}C_{\frac{1}{2}}-\frac{\sqrt{2}}{9}C_{\frac{3}{2}} & \frac{8}{9}C_{\frac{1}{2}}+\frac{1}{9}C_{\frac{3}{2}}
\end{array}\right),
\eea
\bea
V_{\frac{3}{2}^{-}}^{C}=\left(\begin{array}{ccc}
\frac{1}{9}C_{\frac{1}{2}}+\frac{8}{9}C_{\frac{3}{2}} & -\frac{1}{3\sqrt{3}}C_{\frac{1}{2}}+\frac{1}{3\sqrt{3}}C_{\frac{3}{2}} & \frac{\sqrt{5}}{9}C_{\frac{1}{2}}-\frac{\sqrt{5}}{9}C_{\frac{3}{2}}\\
-\frac{1}{3\sqrt{3}}C_{\frac{1}{2}}+\frac{1}{3\sqrt{3}}C_{\frac{3}{2}} & \frac{1}{3}C_{\frac{1}{2}}+\frac{2}{3}C_{\frac{3}{2}} & -\frac{1}{3}\sqrt{\frac{5}{3}}C_{\frac{1}{2}}+\frac{1}{3}\sqrt{\frac{5}{3}}C_{\frac{3}{2}}\\
\frac{\sqrt{5}}{9}C_{\frac{1}{2}}-\frac{\sqrt{5}}{9}C_{\frac{3}{2}} & -\frac{1}{3}\sqrt{\frac{5}{3}}C_{\frac{1}{2}}+\frac{1}{3}\sqrt{\frac{5}{3}}C_{\frac{3}{2}} & \frac{5}{9}C_{\frac{1}{2}}+\frac{4}{9}C_{\frac{3}{2}}
\end{array}\right),
\eea
\bea
V_{\frac{5}{2}^{-}}^{C}  =C_{\frac{3}{2}}.
\eea

\section{One-pion-exchange potentials}
Once the one-pion-exchange potentials (OPEs) are included,
the mixing between the $S$-wave and $D$-wave needs to be considered. The quantum numbers of the $\Sigma_c^{(*)}\bar{D}^{(*)}$
systems taken into account in this work are listed in Table~\ref{tab:quantumnumbers}. 
\begin{table}[!thb]
\caption{Coupled channels for both the $S$- and $D$-wave $\Sigma_c^{(\ast)}\bar{D}^{(\ast)}$ systems. The subindices denote the total spin of the two particles.}
{\begin{tabular}{c|c|c}
\hline
$J^P$ & $S$-wave & $D$-wave  \\
\hline
$\bigg(\dfrac12\bigg)^-$ & $\Sigma_c\bar{D}$, $\Sigma_c\bar{D}^\ast$, $\Sigma_c^\ast\bar{D}^\ast$ & $\Sigma_c\bar{D}^\ast$, $\Sigma_c^\ast\bar{D}$, $\left(\Sigma_c^\ast\bar{D}^\ast\right)_\frac32$, $\left(\Sigma_c^\ast\bar{D}^\ast\right)_\frac52$ \\
$\bigg(\dfrac32\bigg)^-$ & $\Sigma_c\bar{D}^\ast$, $\Sigma_c^\ast\bar{D}$, $\Sigma_c^\ast\bar{D}^\ast$ & $\Sigma_c\bar{D}$, $\left(\Sigma_c\bar{D}^\ast\right)_\frac12$, $\left(\Sigma_c\bar{D}^\ast\right)_\frac32$, $\Sigma_c^\ast\bar{D}$, $\left(\Sigma_c^\ast\bar{D}^\ast\right)_\frac12$, $\left(\Sigma_c^\ast\bar{D}^\ast\right)_\frac32$, $\left(\Sigma_c^\ast\bar{D}^\ast\right)_\frac52$ \\
$\bigg(\dfrac52\bigg)^-$ & $\Sigma_c^\ast\bar{D}^\ast$ &  $\Sigma_c\bar{D}$, $\left(\Sigma_c\bar{D}^\ast\right)_\frac12$, $\left(\Sigma_c\bar{D}^\ast\right)_\frac32$, $\Sigma_c^\ast\bar{D}$, $\left(\Sigma_c^\ast\bar{D}^\ast\right)_\frac12$, $\left(\Sigma_c^\ast\bar{D}^\ast\right)_\frac32$, $\left(\Sigma_c^\ast\bar{D}^\ast\right)_\frac52$ \\
\hline
\end{tabular}\label{tab:quantumnumbers}
}
\end{table}

The isospin wave function for the $P_c^+$ is
\bea
P_c^+=-\sqrt{\frac{1}{3}}\Sigma_c^{(*)+}\bar{D}^{(*)0}+\sqrt{\frac{2}{3}}\Sigma_c^{(*)++}D^{(*)-}.
\eea
The corresponding potential can be written as  
\bea\nonumber
V_{\Sigma_c^{(*)}\bar{D}^{(*)}}&=&\frac{1}{3}V_{\Sigma_c^{(*)+}\bar{D}^{(*)0}\to \Sigma_c^{(*)+}\bar{D}^{(*)0}}+\frac{2}{3}V_{\Sigma_c^{(*)++}D^{(*)-}\to \Sigma_c^{(*)++}D^{(*)-}} \nonumber\\
&& -\frac{\sqrt{2}}{3}V_{\Sigma_c^{(*)+}\bar{D}^{(*)0}\to \Sigma_c^{(*)++}D^{(*)-}}-\frac{\sqrt{2}}{3}V_{\Sigma_c^{(*)++}D^{(*)-}\to \Sigma_c^{(*)+}\bar{D}^{(*)0}} \nonumber\\
&=&2 V_{\Sigma_c^{(*)++}D^{(*)-}\to \Sigma_c^{(*)++}D^{(*)-}},
\eea
with
\bea\nonumber
V_{\Sigma_c^{(*)+}\bar{D}^{(*)0}\to \Sigma_c^{(*)+}\bar{D}^{(*)0}}&=&0\\
V_{\Sigma_c^{(*)++}D^{(*)-}\to \Sigma_c^{(*)++}D^{(*)-}}&=&-\frac{1}{\sqrt{2}}V_{\Sigma_c^{(*)+}\bar{D}^{(*)0}\to \Sigma_c^{(*)++}D^{(*)-}}=-\frac{1}{\sqrt{2}}V_{\Sigma_c^{(*)++}D^{(*)-}\to \Sigma_c^{(*)+}\bar{D}^{(*)0}}. \nonumber
\eea

A state $\left|JM,L\Sigma\right\rangle$ with $J$ ($M$) the total angular momentum (its third component), 
$L$ the orbital angular momentum and $\Sigma$ the total spin can be written as
\bea
\left|JM,L\Sigma\right\rangle=\frac{1}{\sqrt{4\pi}}\sum_{\sigma_{1},\sigma_{2},\Sigma_{3},L_{3}}\int d\Omega_{\vec{p}}Y_{L}^{L_{3}}\left(\Omega_{\vec{p}}\right)\langle s_{1}s_{2}\sigma_{1}\sigma_{2} | \Sigma \Sigma_3\rangle \langle L\Sigma  L_{3}\Sigma_{3}|JM\rangle\left|\vec{p}\sigma_{1}\sigma_{2}\right\rangle
\eea
in terms of $\left|\vec{p}\sigma_1\sigma_2\right\rangle$, which is the direct product of the one-particle states
$\left|\vec{p},\sigma_1\right\rangle$ and $\left|-\vec{p},\sigma_2\right\rangle$ with $\sigma_i$ the third component of 
spin $s_i$ for the $i^{\rm th}$ particle in that channel; $\vec{p}$ is the momentum in the c.m. frame.
In general, all the transitions between states with the 
same $J$ need to be included with the transition matrix elements
\bea
T_{L\Sigma,L^{\prime}\Sigma^{\prime}}^{J}=\langle JM,L\Sigma|\hat{T}|JM,L^{\prime}\Sigma^{\prime}\rangle,
\label{eq:transition}
\eea
where $\hat{T}$ is the transition operator. Furthermore, the partial-wave amplitude can be obtained through
\bea\nonumber
T_{L\Sigma,L^{\prime}\Sigma^{\prime}}^{J}&=&\frac{1}{4\pi}\sum_{\sigma_{1}\sigma_{2}\Sigma_{3}L_{3}}\sum_{\sigma_{1}^{\prime}\sigma_{2}^{\prime}\Sigma_{3}^{\prime}L_{3}^{\prime}}
\int d\Omega_{\vec{p}}Y_{L}^{L_{3}}\left(\Omega_{\vec{p}}\right)\langle s_{1}s_{2}\sigma_{1}\sigma_{2} | \Sigma \Sigma_3\rangle \langle L\Sigma  L_{3}\Sigma_{3}| JM\rangle\\
&&\times\int d\Omega_{\vec{p}^{\prime}}Y_{L^{\prime}}^{L_{3}^{\prime}*}\left(\Omega_{\vec{p}^{\prime}}\right)\langle s_{1}^{\prime}s_{2}^{\prime}\sigma_{1}^{\prime}\sigma_{2}^{\prime} | \Sigma^{\prime} \Sigma_3^{\prime}\rangle \langle L^{\prime}\Sigma^{\prime} L_{3}^{\prime}\Sigma_{3}^{\prime}| JM\rangle\left\langle\vec{p}^{\prime}\sigma_{1}^{\prime}\sigma_{2}^{\prime}\right|\hat{T}\left|\vec{p}\sigma_1\sigma_2\right\rangle.
\eea

In the framework of the time-ordered perturbation theory (TOPT), the OPE potential acquires two contributions~[50]. For the scattering process  $12\to 1^\prime 2^\prime$ with $E$ the total energy of the system, and $p$, $\tilde p$ the incoming and outgoing three-momenta, we define
\bea
V_{SS}(E,p,\tilde p)&\equiv & \frac12 \int_{-1}^{1}d\cos\theta \frac{\tilde p^{2}+p^{2}-2\tilde pp\cos\theta}{2E_\pi(p_\pi)}\big[ D_a^\pi(E,p,\tilde p,\theta)+D_b^\pi(E,p,\tilde p,\theta) \big],\\ 
V_{SD}(E,p,\tilde p)&\equiv & \frac12 \int_{-1}^{1}d\cos\theta \frac{4\tilde p^2+p^2-8\tilde pp\cos\theta+3p^2\cos 2\theta }{2E_\pi(p_\pi)}\big[ D_a^\pi(E,p,\tilde p,\theta)+D_b^\pi(E,p,\tilde p,\theta) \big],\\
V_{DS}(E,p,\tilde p)&\equiv & \frac12 \int_{-1}^{1}d\cos\theta \frac{\tilde p^2+4p^2-8\tilde pp\cos\theta+3p^2\cos 2\theta }{2E_\pi(p_\pi)}\big[ D_a^\pi(E,p,\tilde p,\theta)+D_b^\pi(E,p,\tilde p,\theta) \big],\\
V_{DD}^{(c_1,c_2,c_3,c_4)}(E,p,\tilde p)&\equiv & \frac12 \int_{-1}^{1}d\cos\theta \frac{c_1 (\tilde p^2+p^2)-c_2\,\tilde pp\cos\theta+c_3(\tilde p^2+p^2)\cos 2\theta-c_4\,\tilde pp\cos 3\theta }{2E_\pi(p_\pi)} \nonumber\\ 
&& \times\big[ D_a^\pi(E,p,\tilde p,\theta)+D_b^\pi(E,p,\tilde p,\theta) \big],
\eea
where  $\theta$ denotes the angle between the three-momenta $\vec {\tilde p} $ and $\vec p$.
The values for the $c_i(i=1,\ldots,4)$ coefficients are specified below for each value of the total angular momentum $J$. Moreover, 
\bea
D_a^\pi(E,p,\tilde p,\theta)&=&\frac{1}{E_\pi(p_\pi)+E_{1^\prime}(\tilde p)+E_2(p)-E},\nonumber\\
D_b^\pi(E,p,\tilde p,\theta)&=&\frac{1}{E_\pi(p_\pi)+E_{1}(p)+E_{2^\prime}(\tilde p)-E}\nonumber\\
\eea
are the contributions of the two TOPT orderings with $p_\pi=\sqrt{p^2+\tilde p^2-2p\tilde p\cos\theta}$ and $E_i=\sqrt{p_i^2+m_i^2}$. The OPE potentials in the elastic channels listed in Table~\ref{tab:quantumnumbers} for $J^P=\frac12^-$ can be written in the matrix form (where the columns and rows given by the channels listed in order in the table) as
\bea
V_{\frac{1}{2}^{-}}^\text{OPE}=-2\frac{gg_{1}}{f_{\pi}^{2}}\left(\begin{array}{cc}
V_{\frac{1}{2}^{-}}^{SS} & V_{\frac{1}{2}^{-}}^{SD}\\
V_{\frac{1}{2}^{-}}^{DS} & V_{\frac{1}{2}^{-}}^{DD}
\end{array}\right),
\eea
with 
\[
V_{\frac{1}{2}^{-}}^{SS}=\left(\begin{array}{ccc}
0 & \frac{1}{4\sqrt{3}}V_{SS} & -\frac{1}{4\sqrt{6}}V_{SS}\\
\frac{1}{4\sqrt{3}}V_{SS} & -\frac{1}{6}V_{SS} & -\frac{1}{12\sqrt{2}}V_{SS}\\
-\frac{1}{4\sqrt{6}}V_{SS} & -\frac{1}{12\sqrt{2}}V_{SS} & -\frac{5}{24}V_{SS}
\end{array}\right),
\]
\[
V_{\frac{1}{2}^{-}}^{SD}=\left(\begin{array}{cccc}
\frac{1}{8\sqrt{6}}V_{SD} & 0 & \frac{1}{16\sqrt{30}}V_{SD} & \frac{1}{16}\sqrt{\frac{3}{10}}V_{SD}\\
\frac{1}{24\sqrt{2}}V_{SD} & -\frac{1}{16\sqrt{6}}V_{SD} & \frac{1}{12\sqrt{10}}V_{SD} & -\frac{1}{16\sqrt{10}}V_{SD}\\
\frac{1}{96}V_{DS} & \frac{1}{32\sqrt{3}}V_{SD} & -\frac{7}{96\sqrt{5}}V_{SD} & -\frac{1}{16\sqrt{5}}V_{SD}
\end{array}\right),
\]
\[
V_{\frac{1}{2}^{-}}^{DS}=\left(\begin{array}{ccc}
\frac{1}{8\sqrt{6}}V_{DS} & \frac{1}{24\sqrt{2}}V_{DS} & \frac{1}{96}V_{DS}\\
0 & -\frac{1}{16\sqrt{6}}V_{DS} & \frac{1}{32\sqrt{3}}V_{DS}\\
\frac{1}{16\sqrt{30}}V_{DS} & \frac{1}{12\sqrt{10}}V_{DS} & -\frac{7}{96\sqrt{5}}V_{DS}\\
\frac{1}{16}\sqrt{\frac{3}{10}}V_{DS} & -\frac{1}{16\sqrt{10}}V_{DS} & -\frac{1}{16\sqrt{5}}V_{DS}
\end{array}\right),
\]
\[
V_{\frac{1}{2}^{-}}^{DD}=\left(\begin{array}{cccc}
-\frac{1}{48}V_{DD}^{(1,8,3,0)} & \frac{1}{64\sqrt{3}}V_{DD}^{(4,23,12,9)} & -\frac{1}{192\sqrt{5}}V_{DD}^{(8,37,24,27)} & \frac{1}{64\sqrt{5}}V_{DD}^{(2,13,6,3)}\\
\frac{1}{64\sqrt{3}}V_{DD}^{(4,23,12,9)} & 0 & \frac{1}{32\sqrt{15}}V_{DD}^{(1,-1,3,9)} & \frac{1}{64}\sqrt{\frac{3}{5}}V_{DD}^{(2,13,6,3)}\\
-\frac{1}{192\sqrt{5}}V_{DD}^{(8,37,24,27)} & \frac{1}{32\sqrt{15}}V_{DD}^{(1,-1,3,9)} & -\frac{1}{240}V_{DD}^{(13,77,39,27)} & -\frac{1}{320}V_{DD}^{(2,13,6,3)}\\
\frac{1}{64\sqrt{5}}V_{DD}^{(2,13,6,3)} & \frac{1}{64}\sqrt{\frac{3}{5}}V_{DD}^{(2,13,6,3)} & -\frac{1}{320}V_{DD}^{(2,13,6,3)} & -\frac{3}{160}V_{DD}^{(1,9,3,-1)}
\end{array}\right).
\]
The OPE potentials in the elastic channels listed in Table~\ref{tab:quantumnumbers} for $J^P=\frac32^-$ read
\bea
V_{\frac{3}{2}^{-}}^\text{OPE}=-2\frac{gg_{1}}{f_{\pi}^{2}}\left(\begin{array}{cc}
V_{\frac{3}{2}^{-}}^{SS} & V_{\frac{3}{2}^{-}}^{SD}\\
V_{\frac{3}{2}^{-}}^{DS} & V_{\frac{3}{2}^{-}}^{DD}
\end{array}\right),
\eea
where
\[
V_{\frac{3}{2}^{-}}^{SS}=\left(\begin{array}{ccc}
\frac{1}{12}V_{SS} & \frac{1}{8\sqrt{3}}V_{SS} & -\frac{\sqrt{5}}{24}V_{SS}\\
\frac{1}{8\sqrt{3}}V_{SS} & 0 & \frac{1}{8}\sqrt{\frac{5}{3}}V_{SS}\\
-\frac{\sqrt{5}}{24}V_{SS} & \frac{1}{8}\sqrt{\frac{5}{3}}V_{SS} & -\frac{1}{12}V_{SS}
\end{array}\right),
\]
\[
V_{\frac{3}{2}^{-}}^{SD}=\left(\begin{array}{ccccccc}
-\frac{1}{16\sqrt{3}}V_{SD} & -\frac{1}{48}V_{SD} & \frac{1}{24}V_{SD} & -\frac{1}{32\sqrt{3}}V_{SD} & -\frac{1}{96\sqrt{2}}V_{SD} & -\frac{1}{96\sqrt{5}}V_{SD} & \frac{1}{32}\sqrt{\frac{7}{10}}V_{SD}\\
0 & -\frac{1}{32\sqrt{3}}V_{SD} & -\frac{1}{32\sqrt{3}}V_{SD} & 0 & \frac{1}{32\sqrt{6}}V_{SD} & \frac{1}{8\sqrt{15}}V_{SD} & \frac{1}{32}\sqrt{\frac{21}{10}}V_{SD}\\
-\frac{1}{32\sqrt{15}}V_{SD} & -\frac{1}{24\sqrt{5}}V_{SD} & -\frac{1}{96\sqrt{5}}V_{SD} & \frac{1}{8\sqrt{15}}V_{SD} & \frac{7}{96\sqrt{10}}V_{SD} & \frac{1}{30}V_{SD} & -\frac{1}{160}\sqrt{\frac{7}{2}}V_{SD}
\end{array}\right),
\]
\[
V_{\frac{3}{2}^{-}}^{DS}=\left(\begin{array}{ccc}
-\frac{1}{16\sqrt{3}}V_{DS} & 0 & -\frac{1}{32\sqrt{15}}V_{DS}\\
-\frac{1}{48}V_{DS} & -\frac{1}{32\sqrt{3}}V_{DS} & -\frac{1}{24\sqrt{5}}V_{DS}\\
\frac{1}{24}V_{DS} & -\frac{1}{32\sqrt{3}}V_{DS} & -\frac{1}{96\sqrt{5}}V_{DS}\\
-\frac{1}{32\sqrt{3}}V_{DS} & 0 & \frac{1}{8\sqrt{15}}V_{DS}\\
-\frac{1}{96\sqrt{2}}V_{DS} & \frac{1}{32\sqrt{6}}V_{DS} & \frac{7}{96\sqrt{10}}V_{DS}\\
-\frac{1}{96\sqrt{5}}V_{DS} & \frac{1}{8\sqrt{15}}V_{DS} & \frac{1}{30}V_{DS}\\
\frac{1}{32}\sqrt{\frac{7}{10}}V_{DS} & \frac{1}{32}\sqrt{\frac{21}{10}}V_{DS} & -\frac{1}{160}\sqrt{\frac{7}{2}}V_{DS}
\end{array}\right),
\]
%
\[
{\setstretch{1.35}
V_{\frac{3}{2}^{-}}^{DD}=\left(\begin{array}{ccccccc}
0 & \frac{V_{DD}^{(1,5,3,3)} }{16\sqrt{3}}& \frac{V_{DD}^{(2,13,6,3)}}{32\sqrt{3}} & 0 & -\frac{V_{DD}^{(1,5,3,3)}}{16\sqrt{6}} & \frac{V_{DD}^{(2,13,6,3)}}{64\sqrt{15}} & -\sqrt{\frac{6}{35}}\frac{V_{DD}^{(2,13,6,3)}}{64}\\
\frac{V_{DD}^{(1,5,3,3)}}{16\sqrt{3}} & -\frac{V_{DD}^{(1,5,3,3)}}{24} & \frac{V_{DD}^{(2,13,6,3)}}{96} & \frac{V_{DD}^{(2,13,6,3)}}{64\sqrt{3}} & -\frac{V_{DD}^{(1,5,3,3)}}{48\sqrt{2}} & \frac{V_{DD}^{(2,13,6,3)}}{48\sqrt{5}} & \frac{V_{DD}^{(2,13,6,3)}}{32\sqrt{70}}\\
\frac{V_{DD}^{(2,13,6,3)}}{32\sqrt{3}} & \frac{V_{DD}^{(2,13,6,3)}}{96} & \frac{V_{DD}^{(1,5,3,3)}}{48} & \frac{V_{DD}^{(1,5,3,3)}}{32\sqrt{3}} & \frac{V_{DD}^{(2,13,6,3)}}{192\sqrt{2}} & -\frac{\sqrt{5}}{96}V_{DD}^{(1,5,3,3)} & \sqrt{\frac{5}{14}}\frac{V_{DD}^{(2,13,6,3)}}{64}\\
0 & \frac{V_{DD}^{(2,13,6,3)}}{64\sqrt{3}} & \frac{V_{DD}^{(1,5,3,3)}}{32\sqrt{3}} & 0 & -\frac{V_{DD}^{(2,13,6,3)}}{64\sqrt{6}} & \frac{\sqrt{\frac{5}{3}}V_{DD}^{(1,5,3,3)}}{32} & \sqrt{\frac{15}{14}}\frac{V_{DD}^{(2,13,6,3)}}{64}\\
-\frac{V_{DD}^{(1,5,3,3)}}{16\sqrt{6}} & -\frac{V_{DD}^{(1,5,3,3)}}{48\sqrt{2}} & \frac{V_{DD}^{(2,13,6,3)}}{192\sqrt{2}} & -\frac{V_{DD}^{(2,13,6,3)}}{64\sqrt{6}} & -\frac{5}{96}V_{DD}^{(1,5,3,3)} & -\frac{7V_{DD}^{(2,13,6,3)}}{192\sqrt{10}} & \frac{V_{DD}^{(2,13,6,3)}}{32\sqrt{35}}\\
\frac{V_{DD}^{(2,13,6,3)}}{64\sqrt{15}} & \frac{V_{DD}^{(2,13,6,3)}}{48\sqrt{5}} & -\frac{\sqrt{5}}{96}V_{DD}^{(1,5,3,3)} & \sqrt{\frac{5}{3}}\frac{V_{DD}^{(1,5,3,3)}}{32} & -\frac{7V_{DD}^{(2,13,6,3)}}{192\sqrt{10}} & -\frac{V_{DD}^{(1,5,3,3)}}{48} & -\frac{V_{DD}^{(2,13,6,3)}}{64\sqrt{14}}\\
-\sqrt{\frac{6}{35}}\frac{V_{DD}^{(2,13,6,3)}}{64} & \frac{V_{DD}^{(2,13,6,3)}}{32\sqrt{70}} & \sqrt{\frac{5}{14}}\frac{V_{DD}^{(2,13,6,3)}}{64} & \sqrt{\frac{15}{14}}\frac{V_{DD}^{(2,13,6,3)}}{64} & \frac{V_{DD}^{(2,13,6,3)}}{32\sqrt{35}} & -\frac{V_{DD}^{(2,13,6,3)}}{64\sqrt{14}} & \frac{3}{224}V_{DD}^{(1,3,3,5)}
\end{array}\right).
}
\]
The OPE potentials in the elastic channels in order listed in Table~\ref{tab:quantumnumbers} for $J^P=\frac52^-$ read
\bea
V_{\frac{5}{2}^{-}}^\text{OPE}=-2\frac{gg_{1}}{f_{\pi}^{2}}\left(\begin{array}{cc}
V_{\frac{5}{2}^{-}}^{SS} & V_{\frac{5}{2}^{-}}^{SD}\\
V_{\frac{5}{2}^{-}}^{DS} & V_{\frac{5}{^{2}}^{-}}^{DD}
\end{array}\right),
\eea
where
\[
V_{\frac{5}{2}^{-}}^{SS}=\frac{1}{8}V_{SS},
\]

\[
V_{\frac{5}{2}^{-}}^{SD}=\left(\begin{array}{ccccccc}
-\frac{1}{16\sqrt{10}}V_{SD} & -\frac{1}{32}\sqrt{\frac{2}{15}}V_{SD} & -\frac{1}{32}\sqrt{\frac{7}{15}}V_{SD} & \frac{1}{32}\sqrt{\frac{7}{5}}V_{SD} & -\frac{1}{16\sqrt{15}}V_{SD} & \frac{1}{160}\sqrt{\frac{7}{3}}V_{SD} & \frac{1}{40}\sqrt{\frac{7}{2}}V_{SD}\end{array}\right),
\]

\[
V_{\frac{5}{2}^{-}}^{DS}=\left(\begin{array}{ccccccc}
-\frac{1}{16\sqrt{10}}V_{DS} & -\frac{1}{32}\sqrt{\frac{2}{15}}V_{DS} & -\frac{1}{32}\sqrt{\frac{7}{15}}V_{DS} & \frac{1}{32}\sqrt{\frac{7}{5}}V_{DS} & -\frac{1}{16\sqrt{15}}V_{DS} & \frac{1}{160}\sqrt{\frac{7}{3}}V_{DS} & \frac{1}{40}\sqrt{\frac{7}{2}}V_{DS}\end{array}\right)^{\mathrm{T}},
\]
\[
{\setstretch{1.35}
V_{\frac{5}{2}^{-}}^{DD}=\left(\begin{array}{ccccccc}
0 & \frac{V_{DD}^{(1,5,3,3)} }{16\sqrt{3}}& -\frac{V_{DD}^{(2,13,6,3)} }{16\sqrt{42}}& 0 & -\frac{V_{DD}^{(1,5,3,3)} }{16\sqrt{6}}& -\frac{V_{DD}^{(2,13,6,3)} }{32\sqrt{210}}& -\frac{V_{DD}^{(2,13,6,3)}}{16\sqrt{35}}\\
\frac{V_{DD}^{(1,5,3,3)} }{16\sqrt{3}}& -\frac{V_{DD}^{(1,5,3,3)}}{24} & -\frac{V_{DD}^{(2,13,6,3)}}{48\sqrt{14}} & -\frac{V_{DD}^{(2,13,6,3)}}{32\sqrt{42}} & -\frac{V_{DD}^{(1,5,3,3)}}{48\sqrt{2}} & -\frac{V_{DD}^{(2,13,6,3)}}{24\sqrt{70}} & \frac{V_{DD}^{(2,13,6,3)}}{16\sqrt{105}}\\
-\frac{V_{DD}^{(2,13,6,3)}}{16\sqrt{42}} & -\frac{V_{DD}^{(2,13,6,3)}}{48\sqrt{14}} & \frac{V_{DD}^{(17,100,51,36)}}{336} & \frac{V_{DD}^{(4,5,12,27)}}{448\sqrt{3}} & -\frac{V_{DD}^{(2,13,6,3)}}{192\sqrt{7}} & -\frac{\sqrt{5}V_{DD}^{(16,83,48,45)}}{1344} & \sqrt{\frac{5}{6}}\frac{V_{DD}^{(2,13,6,3)}}{224}\\
0 & -\frac{V_{DD}^{(2,13,6,3)}}{32\sqrt{42}} & \frac{V_{DD}^{(4,5,12,27)}}{448\sqrt{3}} & 0 & \frac{V_{DD}^{(2,13,6,3)}}{64\sqrt{21}} & \sqrt{\frac{5}{3}}\frac{V_{DD}^{(11,61,33,27)}}{224} & \frac{\sqrt{10}}{448}V_{DD}^{(2,13,6,3)}\\
-\frac{V_{DD}^{(1,5,3,3)}}{16\sqrt{6}} & -\frac{V_{DD}^{(1,5,3,3)}}{48\sqrt{2}} & -\frac{V_{DD}^{(2,13,6,3)}}{192\sqrt{7}} & \frac{V_{DD}^{(2,13,6,3)}}{64\sqrt{21}} & -\frac{5}{96}V_{DD}^{(1,5,3,3)} & \sqrt{\frac{7}{5}}\frac{V_{DD}^{(2,13,6,3)}}{192} & \frac{V_{DD}^{(2,13,6,3)}}{8\sqrt{210}}\\
-\frac{V_{DD}^{(2,13,6,3)}}{32\sqrt{210}} & -\frac{V_{DD}^{(2,13,6,3)}}{24\sqrt{70}} & -\frac{\sqrt{5}}{1344}V_{DD}^{(16,83,48,45)} & \sqrt{\frac{5}{3}}\frac{V_{DD}^{(11,61,33,27)}}{224} & \sqrt{\frac{7}{5}}\frac{V_{DD}^{(2,13,6,3)}}{192} & \frac{V_{DD}^{(1,17,3,-9)}}{336} & -\frac{V_{DD}^{(2,13,6,3)}}{224\sqrt{6}}\\
-\frac{V_{DD}^{(2,13,6,3)}}{16\sqrt{35}} & \frac{V_{DD}^{(2,13,6,3)}}{16\sqrt{105}} & \sqrt{\frac{5}{6}}\frac{V_{DD}^{(2,13,6,3)}}{224} & \frac{\sqrt{10}}{448}V_{DD}^{(2,13,6,3)} & \frac{V_{DD}^{(2,13,6,3)}}{8\sqrt{210}} & -\frac{V_{DD}^{(2,13,6,3)}}{224\sqrt{6}} & \frac{V_{DD}^{(11,61,33,27)}}{224}
\end{array}\right).
}
\]

\section{Weak production amplitudes}

Since the energy region of interest is close the $\sigh$ thresholds, we only consider the elastic $\Sigma_c^{(\ast)}\bar{D}^{(\ast)}$ channels produced in an $S$-wave. We parametrize the weak production of the $\left|s_Q\otimes j_{\ell}\right\rangle$ states with the total angular
momentum $J$, $\mathcal{F}^J_{s_Q,j_\ell} = \left\langle \Lambda_b \right|\hat{\mathcal{H}}_W\left|K^-(s_Q\otimes j_{\ell})^J\right\rangle$, as constants,
where $(s_Q\otimes j_{\ell})^J$ refers to the state with $J$ in the $\left|s_Q\otimes j_{\ell}\right\rangle$ basis. Specifically, the production contact terms for the $J=\frac12$ channels are parametrized as
\begin{eqnarray}
   P^\frac{1}{2}_1&=&\frac{1}{2}\mathcal{F}^\frac{1}{2}_{0,\frac{1}{2}}-\frac{1}{2\sqrt{3}}\mathcal{F}^\frac{1}{2}_{1,\frac{1}{2}}+\sqrt{\frac{2}{3}}\mathcal{F}^\frac{1}{2}_{1,\frac{3}{2}},\\
   P^\frac{1}{2}_2&=&-\frac{1}{2\sqrt{3}}\mathcal{F}^\frac{1}{2}_{0,\frac{1}{2}}+\frac{5}{6}\mathcal{F}_{1,\frac{1}{2}}+\frac{\sqrt{2}}{3}\mathcal{F}^\frac{1}{2}_{1,\frac{3}{2}},\\
   P^\frac{1}{2}_3&=&\sqrt{\frac{2}{3}}\mathcal{F}^\frac{1}{2}_{0,\frac{1}{2}}+\frac{\sqrt{2}}{3}\mathcal{F}^\frac{1}{2}_{1,\frac{1}{2}}-\frac{1}{3}\mathcal{F}^\frac{1}{2}_{1,\frac{3}{2}},
  \end{eqnarray}
where the indices $i=1,2,3$ in $P_i^\frac12$ represent the weak production for the $S$-wave $\Sigma_c\bar{D}$, $\Sigma_c\bar{D}^\ast$ and $\Sigma_c^\ast\bar{D}^\ast$, respectively. Analogously, the weak production for the $S$-wave $\Sigma_c\bar{D}^*$, $\Sigma_c^*\bar{D}$ and $\Sigma_c^*\bar{D}^*$ for $J=\frac32$ are written as
\begin{eqnarray}
   P^\frac{3}{2}_1&=&-\frac{1}{\sqrt{3}}\mathcal{F}^\frac{3}{2}_{0,\frac{3}{2}}+\frac{1}{3}\mathcal{F}^\frac{3}{2}_{1,\frac{1}{2}}+\frac{\sqrt{5}}{3}\mathcal{F}^\frac{3}{2}_{1,\frac{3}{2},}\\
   P^\frac{3}{2}_2&=&\frac{1}{2}\mathcal{F}^\frac{3}{2}_{0,\frac{3}{2}}-\frac{1}{\sqrt{3}}\mathcal{F}^\frac{3}{2}_{1,\frac{1}{2}}+\frac{1}{2}\sqrt{\frac{5}{3}}\mathcal{F}^\frac{3}{2}_{1,\frac{3}{2}},\\
   P^\frac{3}{2}_3&=&\frac{1}{2}\sqrt{\frac{5}{3}}\mathcal{F}^\frac{3}{2}_{0,\frac{3}{2}}+\frac{\sqrt{5}}{3}\mathcal{F}^\frac{3}{2}_{1,\frac{1}{2}}+\frac{1}{6}\mathcal{F}^\frac{3}{2}_{1,\frac{3}{2}},
\end{eqnarray}
in order. The $J=\frac52$ production for the $S$-wave $\Sigma_c^*\bar{D}^*$ is
\bea
P^\frac{5}{2}=\mathcal{F}^\frac52_{1,\frac{3}{2}}.
\eea

\section{Additional fits to the ${J/\psi p}$ invariant mass distributions}

To check the stability of the results and the significance of the the narrow $P_c(4380)$ structure, in addition to the fits reported in the main text, we also fit to the inclusive and $m_{Kp}> 1.9$ GeV ${J/\psi p}$ distributions reported by LHCb~[17]. As pointed out in the LHCb analysis,
in the latter case, the cut employed to constraint  
the energy range of $m_{Kp}$ removes over $80\%$ of the $\Lambda^\ast$-resonance  contributions. The signal amplitudes and the background used in the fits are evaluated as described in the main article, see Eqs. (12)
and (14), respectively, and  the discussion around them.  
The results of the fits are displayed in Fig.~\ref{fig:fitsmkkp}. These two data sets can also be described well within the molecular approach, as formulated in the main article, 
 and show a signal of the narrow $P_c(4380)$, too, though the peak in the inclusive case (right panel) is a bit less prominent.
 The   poles and residues for all $P_c$ states extracted from additional fits  agree  within errors with   the ones shown in Table I in the main article. 
\begin{figure*}[!htb]
 \centering
  \includegraphics[width=\textwidth]{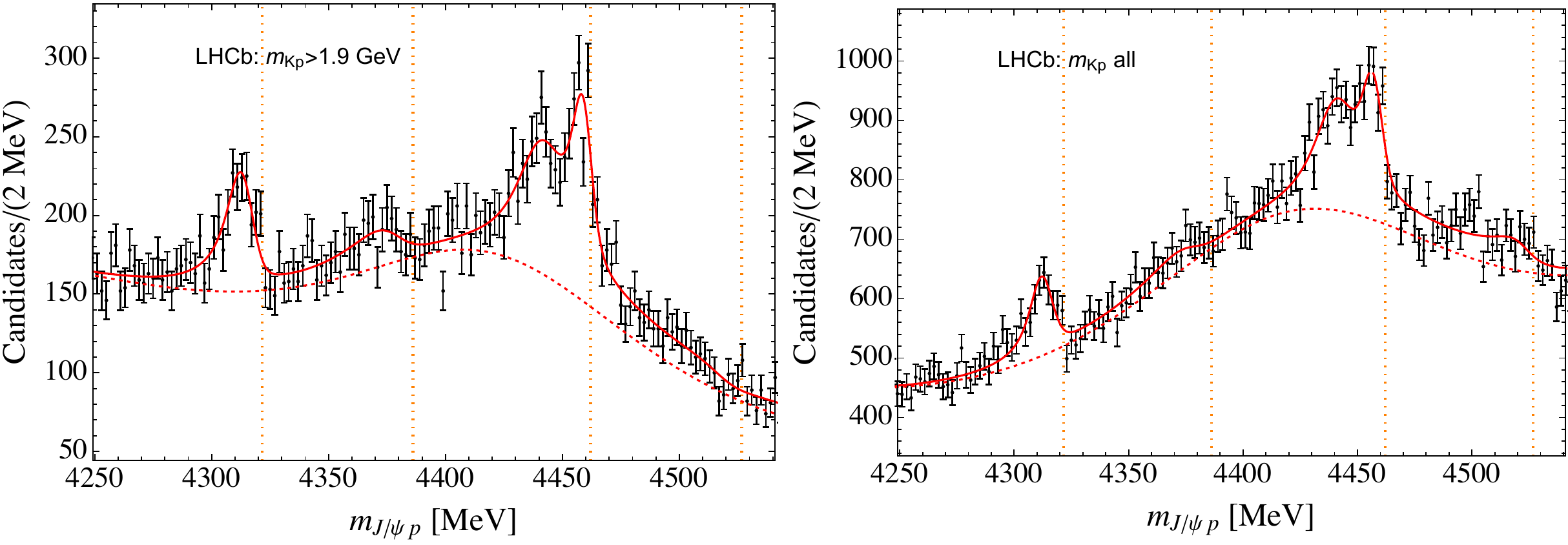}
  \caption{Left panel: the fitted invariant mass distributions versus the experimental data~[17] with the $m_{Kp}>1.9~\mathrm{GeV}$ cut.
   Right panel: the fitted invariant mass distributions versus the inclusive experimental data~[17] (without cuts and weighting).
    }
  \label{fig:fitsmkkp}
\end{figure*}

\section{Formalism}

The product of the effective couplings $g_\alpha g_\gamma$ of a given  $P_c$ state to channels $\alpha$ and $\gamma$ of the $\sigh$ coupled channels is obtained from the residue of the scattering amplitude $T^J_{\alpha \gamma}(E)$ at the pole of that $P_c$, i.e.,
\bea
g_\alpha g_\gamma=\lim_{E\to E_\text{pole}}(E^2-E_\text{pole}^2)T^J_{\alpha \gamma}(E).
\eea
Here the $T$-matrix of the $\sigh\to \sigh$ scattering is given by 
\bea
T_{\alpha\gamma} ^J(E,p, \tilde p)= V_{\alpha\gamma}^J(E,p, \tilde p)-\sum_c\int\frac{d^3\vec{q}}{(2\pi)^3}V_{\alpha\beta}^J(E,p,q)G_\beta(E,q)T_{\beta\gamma}^J(E,q, \tilde p).
\eea
The production strengths $P_\text{pole}$ for a resonance $P_c$ is then defined as 
\bea
P_\text{pole} g_\alpha = \lim_{E\to E_\text{pole}}(E^2-E_\text{pole}^2)U_\alpha^J(E),
\eea
where $U_\alpha^J(E)$ represents the on-shell elastic production amplitude $U_\alpha^J(E,p)$. To account for the widths of $\Sigma_c^{(\ast)}$ of 1.86 MeV (15 MeV) which play an important role in giving the correct widths of resonances, we use a complex mass $m-i\Gamma/2$ in the two-body propagator, i.e.,
\bea
G_\beta (E,q) = \frac{2\mu_\beta}{q^2-p^2_\beta-i\epsilon }, \quad p_\beta^2 =2\mu_\beta(E-m_\text{th}^\beta + i\Gamma^\beta/2 ),
\eea
with $\mu_\beta$, $m_\text{th}^\beta$ and $\Gamma^\beta$ the reduced mass, threshold mass and the width of $\Sigma_c^{(\ast)}$ of the $\beta^\text{th}$ elastic channel.

\end{onecolumngrid}

\end{document}